# The Effect of Ga$^+$ ion beam irradiation on the optical properties of MAPbBr$_3$ single crystals

*Vsevolod I. Yudin, Boris V. Stroganov, Vyacheslav A. Lovtsyus, Olga A. Lozhkina, Anna A. Murashkina, Alexei V. Emeline, Yury V. Kapitonov*

Saint-Petersburg State University, ul. Ulyanovskaya 1, Saint-Petersburg 198504, Russia

**Corresponding Author**

*yury.kapitonov@spbu.ru

The effect of 30 keV Ga$^+$ ion beam irradiation on the low-temperature photoluminescence from a single crystal of the hybrid organic-inorganic perovskite MAPbBr$_3$ is presented for the first time. The irradiation leads to the suppression of the crystal's excitonic luminescence starting from the dose of $10^{14}$/cm$^2$. However, its integral luminescence does not drop up to the doses of two orders of magnitude higher. Further increase of the irradiation dose starts to affect the integral luminescence and causes the crystal to sputter. The spatial resolution of modification by the ion beam found in the experiment is no worse than 50 μm. The theoretical resolution limit estimated by Monte-Carlo modeling using SRIM is three orders of magnitude lower. This work paves the way for the design of devices based on the spatial modulation of perovskite's excitonic properties or direct focused ion beam writing.

**TOC GRAPHICS**

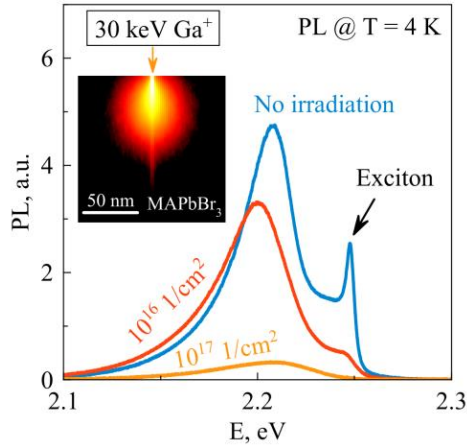

In recent years halide perovskites draw attention of scientific community as a good candidate for solar cell absorber material reaching 20% conversion efficiency[1]. They also exhibit unique optical properties suitable for optoelectronics, laser technology, X-ray and Gamma-ray detection, optical imaging etc. For many uses perovskites in single crystalline form are needed since in such form they have the lowest inhomogeneous broadening of the exciton resonance. Such crystals of good quality could be grown using simple solution-processing techniques[2].

In order to make devices based on halide perovskite single crystals it is important to propose a method of spatial modulation of their optical properties, the best way at subwavelength dimensions. Recently it was shown that ion-beam irradiation of A3B5 quantum wells could lead to the modulation of only the desired exciton properties at ranges suitable for making diffractive optical elements[3]. In this work we study the effect of 30 keV Ga+ ion beam irradiation on the optical properties of the MAPbBr3 single crystal revealing possible route for defect engineering in this material.

Solution grown MAPbBr$_3$ single crystal was irradiated by uniform 300x300 mkm squares using 30 keV Ga$^+$ ion beam (see Supplementary information for growth, sample characterization and irradiation details) with irradiation doses ranging from $10^{10}$ to $10^{17}$ 1/cm$^2$ (Fig.S2,a). SEM image (Fig.S1) shows that sample surface became rough starting from $10^{16}$ 1/cm$^2$ dose designating the transition to the surface sputtering regime. Same effect was observed for white light scattering in an optical microscope (Fig.S1). Square with $10^{17}$ 1/cm$^2$ dose appears black most probably indicating material decomposition accompanying sputtering. At lower doses the main effect of irradiation is the generation of defects in the material.

In order to address exciton properties the sample was cooled down to 4 K. Fig.1 shows PL spectra from non-irradiated and irradiated areas. One could see that in the pristine crystal PL shows spectrally-narrow (HWHM = 2.5 meV) exciton luminescence centered at $E_0$ = 2.247 eV and broad defect-related recombination at lower energies. Ion beam irradiation leads to the suppression of the exciton signal, and for bigger doses – suppression of the integral PL signal.

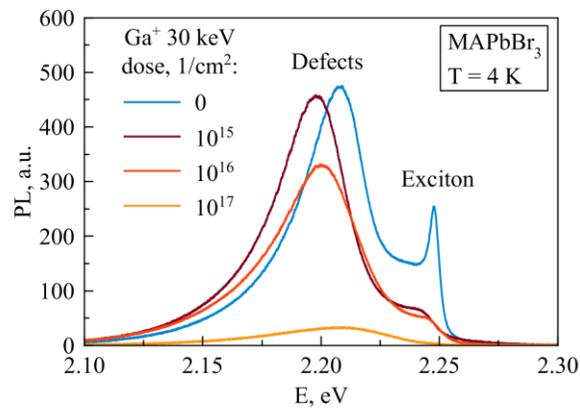

**Figure 1.** PL spectra at 4 K from non-irradiated (blue) and ion-irradiated areas. Spectrally narrow exciton emission and broad defect-related luminescence could be distinguished.

In order to study these effects in more details PL mapping was carried out. Fig. 2, a shows scan of PL spectra along X axis through squares with different ion irradiation doses. Despite the non-uniformity of the sample one could see that exciton emission energy remains unchanged through the sample, and only its intensity is affected by the ion irradiation. Fig. 2,b shows spectrally-integrated PL signal which is determined mainly by the defect-related PL signal. Stability of exciton resonance energy over the sample makes it possible to carry out 2D PL map of the whole sample registered at exciton resonance (Fig. S2,a). Fig.S2, b shows integral PL intensity map.

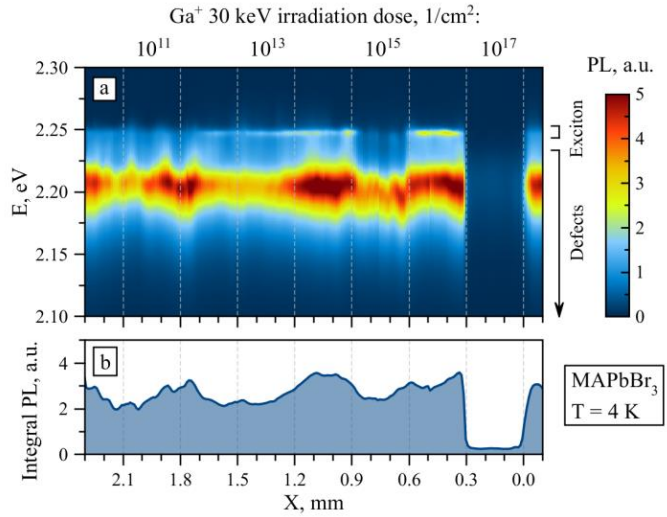

**Figure 2.** (a) Spatial scan of PL spectra along X axis at Y = 5.45 mm through squares with different ion irradiation doses (indicated above). Exciton and defect-related PL signal could be spectrally separated. (b) Spectrally-integrated PL from (a).

Observed effects could be explained as follows. At doses below $10^{14}$ 1/cm² no ion irradiation effect is observed. It means that ion beam induced defects density is well below the background defect density. At $10^{14}$ and $10^{15}$ 1/cm$^2$ doses exciton signal is gradually suppressed due to the inelastic scattering of excitons on high density ion-induced defects. Nevertheless integral PL intensity remains constant indicating that radiation defects do not effectively participate in non-radiative recombination of carriers. At even higher doses sample sputtering begins with prominent decrease of detected integral PL. This decrease is induced by surface layer degradation (which affects both pump light absorption and percent of PL signal escaping the sample), and possibly also by non-radiative recombination at defects. These findings are summarized in Fig.3. Dose dependency of the exciton PL shows exponential decay with dose with the characteristic decay dose $D_0 = 6 \cdot 10^{15}$ 1/cm$^2$.

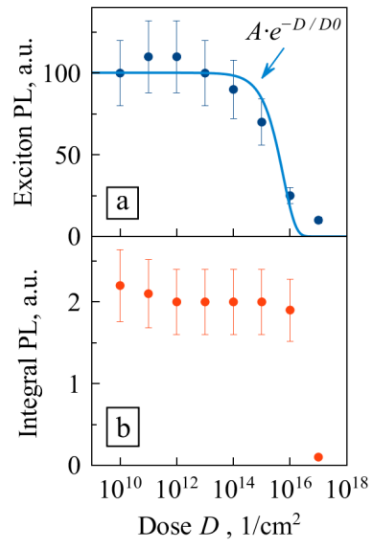

**Figure 3.** Experimental dose dependency of the exciton PL intensity (a) and integral PL (b) averaged over exposed areas (dots). Exciton PL fit by exponential decay (solid line).

Spatial resolution of PL maps was limited by 50 mkm laser spot diameter. Edges of PL profiles of exposed areas shows the same slope (Fig.S3), so the spatial resolution of the ion beam modification of the excitonic resonance is no worse that this value. In order to estimate theoretical limit of the spatial resolution a modelling of defects generation was carried out using Monte-Carlo method. For focused 30 keV Ga$^+$ ion beam irradiation and MAPbBr$_3$ target most vacancies are generated within the hemisphere of 50 nm radius. For lower ion beam energies this volume could be reduced further down to 10 nm hemisphere at 1 keV. In this case ion beam spot size on the target will be the main resolution limitation factor. Such a small interaction volume in contrast to the irradiation of Si and GaAs targets could be explained by stopping action of Pb ions in MAPbBr$_3$.

In this paper we have studied effect of 30 keV Ga$^+$ ion beam irradiation on optical properties of a MAPbBr$_3$ single crystal. Low temperature photoluminescence study of uniformly irradiated areas reveals that starting from $10^{14}$ 1/cm$^2$ ion dose exciton PL intensity is gradually suppressed without change of the integral PL intensity. Only at doses > $10^{16}$ 1/cm$^2$ integral intensity starts to drop together with the sample sputtering and possibly decomposition. This proves high radiation stability of this material with possibility to modulate resonant excitonic properties using moderate ion irradiation doses. Exciton resonance energy remains nearly unchanged through the irradiation which means that the main effect of the irradiation-generated effects is non-elastic scattering of excitons. Monte-Carlo modelling of the defect generation in the sample by ion beam irradiation shows very small interaction volumes definitely subwavelength in size. This observation has two main conclusions: 1. Ion beam irradiation with doses < $10^{16}$ 1/cm$^2$ could be used for creation of active diffractive optical elements, photonic crystals or metamaterials based on the exciton resonance modulation by introduction of additional structural defects. 2. Nanoscale-sized structures retaining resonant excitonic properties could be created from perovskite single crystals by direct focused-ion-beam (FIB) patterning followed by < 100 nm thin damaged layer removal by less destructive methods such as liquid etching or low-energy ion polishing.

**Supporting Information**.

The following files are available free of charge.

Synthesis procedure and structure characterization of MAPbBr3 and experimental details of PL measurements. (PDF)

**Notes**

The authors declare no competing financial interests.

ACKNOWLEDGMENTS

   Synthesis of samples was performed within the Project "Establishment of the Laboratory Photoactive Nanocomposite Materials" No. 14.Z50.31.0016 supported by a Mega-grant of the Government of the Russian Federation. Y.V.K. and O.A.L. acknowledge the Russian Science Foundation (Grant 17-72-10070) for financial support of ion beam irradiation and optical study. The work was carried out using equipment of the resource centers "Nanophotonics" and "Geomodel" of Saint-Petersburg State University.

# REFERENCES

(1) Even, J.; Pedesseau, L.; Katan, C.; Kepenekian, M.; Lauret, J.-S.; Sapori, D.; Deleporte, E. Solid-State Physics Perspective on Hybrid Perovskite Semiconductor. *J. Phys. Chem. C* **2015**, *119*, 10161−10177.

(2) O.A. Lozhkina, V.I. Yudin, A.A. Murashkina, V.V. Shilovskikh, V.G. Davydov, R. Kevorkyants, A.V. Emeline, Yu.V. Kapitonov, and D.W. Bahnemann, Low Inhomogeneous Broadening of Excitonic Resonance in MAPbBr3 Single Crystals, *J. Phys. Chem. Lett.* **2018**, *9*, 302−305.

(3) Yu. V. Kapitonov, P. Yu. Shapochkin, L. Yu. Beliaev, Yu. V. Petrov, Yu. P. Efimov, S. A. Eliseev, V. A. Lovtcius, V. V. Petrov, and V. V. Ovsyankin, Ion-beam-assisted spatial modulation of inhomogeneous broadening of a quantum well resonance: excitonic diffraction grating, *Optics Letters* **2016**, *41*, No.1 104.

**SUPPLEMENTARY MATERIAL**

1. Growth and characterization

MAPbBr$_3$ single crystals were synthesized by the inverse temperature crystallization method from dimethyl sulfoxide (DMSO) solution. The solution of 1 mol/l of MABr and 1 mol/l of PbBr2 in DMSO was prepared, filtered, and kept in autoclave at 70º C for 4 hours. The single crystals were isolated. Structural characterization of the solution-grown single crystals was carried out using Scanning Electron Microscope (SEM) Hitachi S-3400N by Energy-Dispersive X-ray (EDX) spectrometer Oxford Instruments X-Max 20 and Electron Backscatter Diffraction (EBSD) using AZtecHKL Channel 5.

2. Ion beam irradiation

MAPbBr$_3$ single crystal was irradiated by 30 keV Ga$^+$ ion beam using Zeiss Crossbeam 1540XB FIB-SEM. Irradiated areas were 300x300 mkm with 300 mkm spacing between them. Irradiated pattern is shown in Fig.S2,a.

3. Optical study

Irradiated sample was cooled down to 4 K by a closed-cycle helium cryostat Montana Cryostation. Sample was illuminated by a cw-laser with wavelength 450 nm focused by a lens in a 50-mkm spot. PL was collected by the same lens, filtered by an yellow filter from scattered laser light and recorded by a home-build spectrometer based on a monochromator and a CCD-array detector. PL mapping was done by an automated scan of the lens along the X-axis by a motorized micrometer stage and manually along the Y-axis by a micrometer translation stage.

4. Monte-Carlo modelling

Monte-Carlo modelling was carried out using TRIM/SRIM software. MAPbBr3 target density was set as 3.582 g/cm$^3$. Modelling was carried out on the 100x100 spatial array with 2 nm step. Ion beam spot at the target surface was set as 1 nm. Secondary cascades were taken into account. For each spatial bin a vacancy generation yield *V* (Vacancies/Angstrom*ion) was calculated. Calculation was carried out for incident ion energies 30, 15, 5 and 1 keV (Fig.S4).

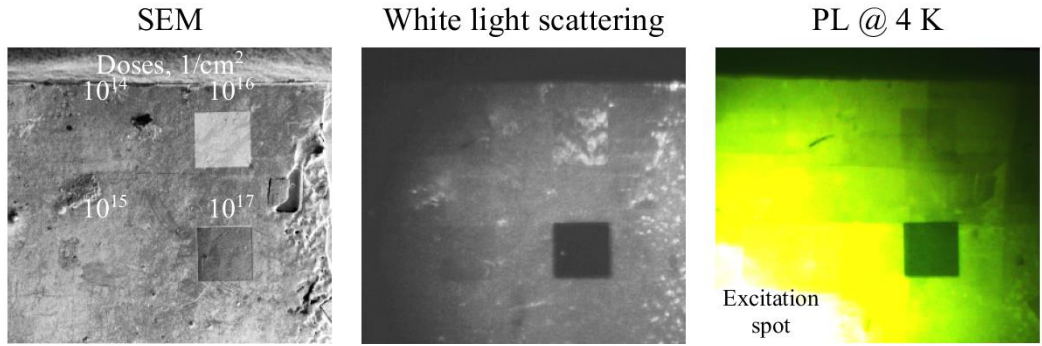

**Figure S1.** (Left) SEM image in secondary electrons mode. (Middle) White light reflection image (color image converted to the grayscale). Note the hardly visible 1015 and 1014 1/cm2 squares. (Right) PL map taken at 4 K with excitation by 450 nm laser and detection by a camera with yellow glass filter removing scattered laser light.

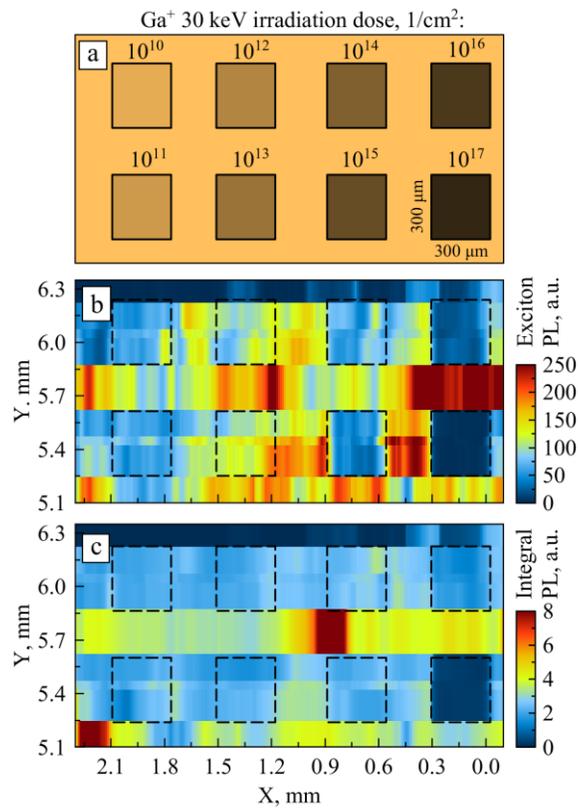

**Figure S2.** (a) Pattern irradiated by ions. 300x300 mkm squares are separated by 300 mkm. (b) PL intensity map for exciton resonance $E_0$ = 2.247 eV (b) and integral PL intensity map (c). All three panels have the same spatial scale. Note the decrease of both resonant and integral PL for $10^{16}$ and $10^{17}$ 1/cm$^2$ doses and decrease of solely resonant PL for $10^{15}$ and $10^{14}$ 1/cm$^2$ doses.

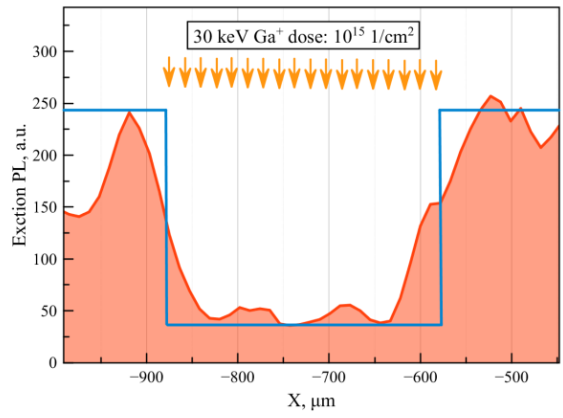

**Figure S3.** Spatial resolution of the ion modification. Red line – exciton PL profile, blue line – irradiated 300 mkm pattern.

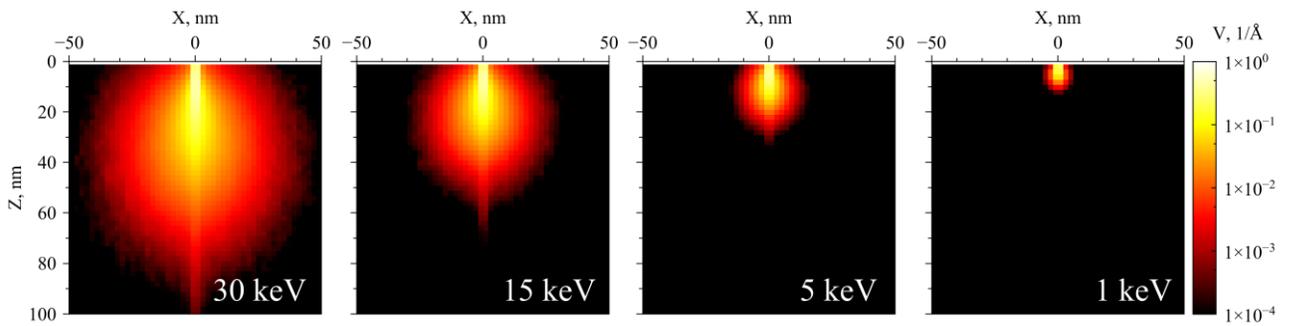

**Figure S4.** Monte-Carlo modelling of the vacancy generation yield $V$ in the case of Ga$^+$ ion irradiation of MAPbBr$_3$ target for ion energies 30, 15, 5 and 1 keV. Ion beam is incident from above at the point X = 0.